\documentclass[10pt]{report}
\usepackage{graphicx}
\usepackage{rotating}
\usepackage[section]{placeins}

\begin{document}

\Large
\centerline{\bf Digging into the surface of the icy dwarf} 

\centerline{\bf planet Eris} 

\vskip 0.5truein

\normalsize
\centerline{M. R. Abernathy} 
\centerline{Dept. Physics \& Astronomy, Northern Arizona Univ, 
Flagstaff, AZ 86011}
\centerline{E-mail: mra66@nau.edu}
\vskip 0.5truein

\centerline{S.  C. Tegler$^1$}
\centerline{Dept. Physics \& Astronomy, Northern Arizona Univ, 
Flagstaff, AZ 86011}
\centerline{E-mail: Stephen.Tegler@nau.edu}
\vskip 0.5truein

\centerline{W. M. Grundy} 
\centerline{Lowell Observatory, 1400 W. Mars Hill Rd., 
Flagstaff, AZ 86001}
\centerline{E-mail: W.Grundy@lowell.edu}
\vskip 0.5truein

\centerline{J. Licandro} \centerline{Instituto de Astrofisica de
Canarias, via Lactea s/n, E38205, La Laguna, Tenerife,
Spain}\centerline{Email: jlicandr@iac.es}
\vskip 0.5truein
\clearpage

\centerline{W. Romanishin$^1$}
\centerline{Dept. Physics \& Astronomy, 
Univ of Oklahoma, Norman, OK 73019}
\centerline{E-mail: wjr@nhn.ou.edu}
\vskip 0.5 truein

\centerline{D. Cornelison}
\centerline{Dept. Physics \& Astronomy, Northern Arizona Univ, 
Flagstaff, AZ 86011}
\centerline{E-mail: David.Cornelison@nau.edu}
\vskip 0.5truein

\centerline{F. Vilas$^1$} \centerline{MMT Observatory, PO Box 210065,
University of Arizona, Tucson, AZ, 85721} \centerline{Email:
fvilas@mmto.org}
\vskip 0.5truein

\vskip 0.7truein

\centerline{Pages: 26; Tables 3; Figures: 5}
\vskip 1.0truein

$^1$ Observer at the MMT Observatory. Observations reported here were
obtained at the MMT Observatory, a joint facility of the University of
Arizona and the Smithsonian Institution.

\vfil\eject

\noindent{Proposed Running Head: \quad Eris\hfil}
\vskip 0.5truein

\noindent{Editorial Correspondence to: \hfil}

\noindent{Dr. Stephen C. Tegler \hfil}

\noindent{Dept Physics \& Astronomy \hfil}

\noindent{Northern Arizona University \hfil}

\noindent{Flagstaff, AZ 86011 \hfil}

\noindent{Phone: 928-523-9382 \hfil}

\noindent{FAX: 928-523-1371 \hfil}

\noindent{E-mail: Stephen.Tegler@nau.edu \hfil}

\vfil\eject


\noindent {\bf Abstract \hfil} 

We describe optical spectroscopic observations of the icy dwarf planet
Eris with the 6.5 meter MMT telescope and the Red Channel
Spectrograph. We report a correlation, that is at the edge of
statistical significance, between blue shift and albedo at maximum
absorption for five methane ice bands. We interpret the correlation as
an increasing dilution of methane ice with another ice component,
probably nitrogen, with increasing depth into the surface.

We suggest a mechanism to explain the apparent increase in nitrogen
with depth.  Specifically, if we are seeing Eris 50 degrees from
pole-on (Brown and Schaller, 2008), the pole we are seeing now at
aphelion was in winter darkness at perihelion. Near perihelion,
sublimation could have built up atmospheric pressure on the sunlit
(summer) hemisphere sufficient to drive winds toward the dark (winter)
hemisphere, where the winds would condense. Because nitrogen is more
volatile and scarcer than methane, it sublimated from the sunlit
hemisphere relatively early in the season, so the early summer
atmosphere was nitrogen rich, and so was the ice deposited on the
winter pole. Later in the season, much of the nitrogen was exhausted
from the summer pole, but there was plenty of methane, which continued
to sublimate. At this point, the atmosphere was more depleted in
nitrogen, as was the ice freezing out on top of the earlier deposited
nitrogen rich ice.

Our increasing nitrogen abundance with depth apparently contradicts
the Licandro et al. (2006) result of a decreasing nitrogen abundance
with depth.  A comparison of observational, data reduction, and
analysis techniques between the two works, suggests the difference
betweeen the two works is real. If so, we may be witnessing the
signature of weather on Eris. The work reported here is intended to
trigger further observational effort by the community.

\smallskip

\bigskip

\noindent{Key Words: Kuiper Belt Objects, Spectroscopy,
Trans-Neptunian Objects \hfil} 
\vfil\eject

\noindent {\bf 1. Introduction \hfil}
\smallskip

Triton, Pluto, Eris (136199 and 2003 UB313), and Makemake (136472 and
2005 FY9) form a natural class of outer Solar System objects for
comparative studies, since their spectra are dominated by strong
methane ice absorption bands (see e.g.  Cruikshank et al., 1993; Owen
et al., 1993; Brown et al., 2005; Licandro et al., 2006). Furthermore,
the same methane ice bands can be used as an important diagnostic tool
in the chemical and physical characterization of icy dwarf planets.

Specifically, it is possible to use the bands to measure the abundance
of methane relative to another ice component, probably nitrogen ice.
Laboratory studies showed that the wavelengths of near-infrared
methane ice absorption bands shift to shorter wavelengths when methane
is diluted by nitrogen ice (Quirico and Schmitt, 1997).  Triton and
Pluto exhibit a nitrogen ice band at 2.15 $\mu$m (Cruikshank et al.,
1993; Owen et al., 1993). Triton's methane ice bands are blue shifted
by 7 nm, suggesting methane is highly diluted by nitrogen ice
(Cruikshank et al., 1993).  Pluto's methane ice band shifts are
smaller, suggesting Pluto's methane occurs in a mixture of diluted and
undiluted phases (Owen et al., 1993).  No one has detected the 2.15
$\mu$m nitrogen ice band in the spectrum of Eris, and its
near-infrared methane bands exhibit even smaller blue shifts which may
be due to the presence of even smaller amounts of nitrogen (Brown et
al., 2005; Dumas et al., 2007).

Second, it is possible to measure the CH$_4$/N$_2$ abundance as a
function of depth into the surface of these objects. The average
penetration depth of a photon at a particular wavelength depends on
the reciprocal of the absorption coefficient at that wavelength,
e.g. photons corresponding to larger absorption coefficients are
absorbed more, preventing them from penetrating as deeply into the
surface. In other words, stronger bands in the spectrum of an icy
dwarf planet probe, on average, shallower into the surface than weaker
bands. So, it is possible to use blue shifts and albedos at maximum
absorption of two or more methane ice bands to measure the trend of
CH$_4$/N$_2$ into the surface of an icy dwarf planet.

Licandro et al. (2006) applied such a technique to Eris. They found
the weaker 729.6 nm band had a shift of 0.1 $\pm$ 0.3 nm and the
stronger 889.7 nm band had a shift of 1.5 $\pm$ 0.3 nm. Their
measurements suggested a decreasing nitrogen abundance with increasing
depth. They speculated that as Eris moved toward aphelion during the
last 200 years, and as its atmosphere cooled, less volatile methane
began to condense out first. As the atmosphere cooled further, it
became more and more nitrogen rich, until the much more volatile
nitrogen could condense out on top of the methane rich ice.

The technique applied by Licandro et al. (2006) is diagnostically
rich. They demonstrated it was possible to put constraints on
atmosphere-surface interactions of icy dwarf planets. So, we set out
to confirm their result by observing additional methane bands in the
spectrum of Eris. Below we describe our blue shift and albedo
measurements and analysis for five methane bands.

\bigskip

\noindent {\bf 2. Observations \hfil}
\smallskip

We obtained optical spectra of Eris on the nights of 2007 September 8
$-$ 11 UT with the 6.5 meter MMT telescope on Mt. Hopkins, Arizona,
the Red Channel Spectrograph, and a new red sensitive, deep depletion
CCD. We used a 1 $\times$ 180 arc sec entrance slit and a 600 g
mm$^{-1}$ grating that provided wavelength coverage of 660.2 $-$ 853.0
nm in first order on the night of 2007 September 8 UT, and 758.7 $-$
951.3 nm in first order on the nights of 2007 September 9, 10, and 11
UT. All spectra had a dispersion of 0.20 nm pixel$^{-1}$, and a
full-width at half-maximum resolution of 0.63 nm.  There were high,
thin cirrus clouds and the seeing was variable, 1 $-$ 2 arc sec on all
nights.  Eris was placed at the center of the slit and the telescope
was tracked at Eris' rate. In Table 1, we present the number of 600
second exposures taken on each night, range of UT times, and range of
airmass values for our Eris observations. Eris had an airmass of 1.25
at transit.

We used the same calibration techniques on Eris spectra as on Makemake
spectra (Tegler et al. 2007; Tegler et al. 2008).  HeNeAr spectra were
taken before and after each set of object spectra to obtain an
accurate wavelength calibration.  Telluric and Fraunhofer lines were
removed from the Eris spectra by dividing each Eris spectrum by the
spectrum of a solar analog star, SA 93$-$101.  The airmass difference
between each Eris spectrum and its corresponding solar analog spectrum
was $<$ 0.1.

Since we saw no significant difference between the individual 600
second exposures of a night, we summed the exposures of each night to
yield four spectra with exposure times of 2 hr 20 min (2007 Sep 8 UT),
2 hr 20 min (2007 Sep 9 UT), 3 hr 00 min (2007 Sep 10 UT), and 2 hr 50
min (2007 Sep 11 UT). In Figure 1, we present our reflectance spectra
of September 9, 10, and 11 UT, and a spectrum of pure methane ice. The
spectra were normalized to a reflectance of 1.0 between 820 and 840
nm. The September 9 and 10 spectra were shifted upward on Figure 1 by
1.0 and 0.5. Clearly, methane ice bands are present in our
spectra.

To assess the uncertainty in our wavelength calibration, we measured
the wavelengths of 16 well-resolved sky emission lines across our
spectra and compared them to the VLT high$-$spectral resolution sky
line atlas of Hanuschik (2003). In Table 2, we present the sky line
measurements. Columns 1 $-$ 4 contain our values, column 5 contains
Hanuschik's values, and columns 6 $-$ 9 contain the differences
between our and Hanuschik's values, all entries are in units of nm.
The average difference and standard deviation of the differences for
each night are given at the bottom of columns 6 $-$ 9. The four
average differences for the four nights are less than $\sim$ 0.03
nm. So, it appears our sky lines, and hence our wavelength
measurements, are accurate to $\sim$ 0.03 nm.

\bigskip

\noindent {\bf 3. Analysis \hfil}
\smallskip

In order to accurately measure methane band shifts in the spectra of
Eris, it is necessary to compare the Eris spectra to a pure methane
ice spectrum. We used Hapke theory to transform laboratory optical
constants of pure methane ice at 30 K (Grundy et al., 2002) into a
spectrum suitable for comparison to Eris spectra by accounting for the
multiple scattering of light within a surface composed of particulate
methane ice (Hapke, 1993).  We used Hapke model parameters of h $=$
0.1, B$_o$ $=$ 0.8, $\overline{\theta}$ $=$ 30$^{\circ}$, P(g) $=$ a
two component Henyey-Greenstein function with 80$\%$ in the forward
scattering lobe and 20$\%$ in the back scattering lobe, and both lobes
had asymmetry parameter a = 0.63.  We fitted each Eris methane band
with two different grain sizes. For example, we fitted the 889.7 nm
band with 1.0 cm (81$\%$ by volume) and 0.8 mm (19$\%$ by volume)
grains.  We explored Hapke parameter space with Monte Carlo techniques
and found a dozen or so of 10,000 Hapke models gave equally good
$\chi^2$ fits for each band. Furthermore, Tegler et al. (2008) found
that relatively small changes to the best Hapke parameters had little
effect on the final shifts. It is important to recognize that our
Hapke parameters do not represent unique fits to the Eris bands;
however, they are plausible values for transparent, pure methane ice
grains and are comparable to Hapke parameters in fits of Pluto spectra
(Grundy and Buie, 2001).

In Figure 2, we present the portions of our MMT spectra in Figure 1
containing the 889.7 nm bands (black lines), and pure methane ice
Hapke models (red lines).  A visual inspection of Figure 2 clearly
shows that the absorption maxima in the Eris spectra are blue shifted
relative to the absorption maxima in the pure methane ice Hapke
spectra.

To quantify the apparent shift, we performed cross-correlation
experiments between the Eris and model spectra in Figure 2.
Specifically, we shifted the Hapke model spectrum in 0.2 nm steps
between $-$3 nm and $+$3 nm, calculated an adjusted $\chi^2$ goodness
of fit, R, for each shift, plotted R vs. shift for each band, and
fitted a parabola to the plot for each band to determine the shift
corresponding to the minimum in R for each band. For the 889.7 nm band
in the September 9, 10, and 11 Eris spectra, we found blue shifts of
0.49 $\pm$ 0.08 nm, 0.51 $\pm$ 0.10 nm, and 0.31 $\pm$ 0.23 nm, i.e. we
found similar shifts for the three nights.

To measure shifts for bands weaker than the 889.7 nm band, it was
essential for us to add the spectra of September 9, 10, and 11. The
similar appearance of the strong 889.7 nm bands in Figure 2, and their
similar shifts encouraged us to co-add the spectra from three
nights. After coadding, the spectrum was again normalized to 1.0
between 820 and 840 nm.

In Figure 3, we present portions of the co-added 2007 September 9, 10,
and 11 UT spectrum exhibiting the 869.1 nm band, 889.7 nm band, and
the 896.8 nm band with the 901.9 nm band (black lines). In addition, we
present a portion of the 2007 September 8 UT spectrum exhibiting the
729.6 nm band (black line). The red lines represent our pure methane
Hapke models. The absorption maxima of the Eris methane bands appear
blue shifted relative to the absorption maxima of the pure methane ice
models. In Table 3, we present the results of our cross correlation
experiments between the Eris bands and the pure methane ice bands in
Figure 3. We point out that even in the co-added spectrum with a total
exposure time of 8 hr 10 min, the signal precision was insufficient
for us to measure shifts for the 786.2 nm, 799.3 nm, 841.5 nm, and
844.2 nm methane bands.

We calculated the uncertainties in the blue shifts as follows.  For
each methane band, the best fit (shifted) model was subtracted from
the corresponding Eris band, giving us the noise in the astronomical
spectrum. We note that the noise in the astronomical spectrum
dominates noise in the Hapke model spectrum. Next, we calculated the
standard deviation of the noise. Then, we applied Gaussian noise with
the same standard deviation to the model spectrum. Next, we applied
our cross correlation technique to the noisy model and the original
model spectra. We generated 1000 different noisy model spectra and
repeated the cross correlation experiment for each noisy model and the
original model spectra.  A histogram of the resulting 1000 shifts was
fit with a Gaussian distribution, the standard deviation of the fit
giving the uncertainty in our shift measurement. We give the
uncertainties in Table 3.

Next, we measured albedos at maximum absorption for each
band. Specifically, normalizing our spectrum to 1.0 between 820 and
840 nm makes our relative reflectance spectrum consistent with albedo
measurements of Stansberry et al. (2008). Therefore, our relative
reflectance spectrum is also an albedo spectrum.

In addition, we estimated the average depths sampled by photons
corresponding to different methane ice bands.  The average penetration
depth of a photon depends not only on its corresponding absorption
coefficient, but also on the amount of scattering, which in turn
depends on particle size, shape, and spacing. We used a Monte Carlo
ray tracing model to explore the trajectories followed by observable
photons at different wavelengths, and thereby estimated the average
depths sampled by photons corresponding to the different methane ice
bands (Grundy and Stansberry, 2000).

In Figure 4, we present a plot of blue shift (which we take as a proxy
for nitrogen dilution) vs. albedo at maximum absorption (horizontal
scale at the bottom of the plot) and average sample depth estimated
from the Monte Carlo model (horizontal scale at the top of the plot)
for the five methane bands in our study (solid squares).

A visual inspection of the solid squares in Figure 4, suggests there
is a correlation between blue shift and albedo. The solid line is a
least square fit to the five points.  In order to estimate the
statistical significance of the apparent correlation in Figure 4, we
computed the Spearman rank correlation coefficient, r, using the
function CORR in the MATLAB programming environment.  The coefficient
takes on values in the range $-$ 1 $<$ r $<$ $+$1, where $-$1
indicates a perfect anti-correlation, 0 indicates no correlation, and
$+$1 indicates a perfect correlation.  We computed r $=$ 0.97 for the
five points in Figure 4. Furthermore, we computed a probability of r
$=$ 0.97 occurring for an uncorrelated sample at 0.03. It appears the
correlation is statistically significant at the 2 sigma level.
\bigskip

\noindent {\bf 4. Atmosphere$-$Surface Interaction Model \hfil}
\smallskip

Is it possible to come up with a mechanism to explain an increasing
nitrogen abundance with increasing depth into the surface of Eris?  We
start with two observational constraints. First, the lack of the
nitrogen ice band at 2.15 $\mu$m in the spectrum of Eris (Brown et
al., 2005; Dumas et al., 2007) suggests the CH$_4$/N$_2$ abundance may
be significantly larger on Eris than on Pluto. Furthermore, Eris has
much smaller blue shifts of the 729.6 nm and 889.7 nm bands (0.70 nm
and 0.46 nm) compared to Pluto (about 3 nm and 1 nm), again suggesting
a larger CH$_4$/N$_2$ abundance on Eris compared to Pluto.  Second,
Eris likely has strong seasonal effects.  Its orbit carries it from a
perihelion distance of 38 AU to an aphelion distance of 97 AU, its
current position. Furthermore, Eris may have a sub-solar latitude of
40$^{\circ}$ at present (Brown and Schaller, 2008).  Hence, we may be
seeing Eris about 50$^{\circ}$ from pole-on, and if so the pole we are
seeing now near aphelion was in permanent winter darkness at
perihelion.  Near perihelion, sublimation could have built up
atmospheric pressure on the sunlit hemisphere (summer pole) sufficient
to drive winds toward the dark hemisphere (winter pole), where the
winds would condense, i.e. the winter pole acted as a cold
trap. Because nitrogen was more volatile and scarcer than methane, it
sublimated from the summer hemisphere relatively early in the season,
so the early summer atmosphere was comparatively nitrogen rich, and so
was the early season ice deposited on the winter pole. Later in the
season, much of the nitrogen was exhausted from the summer pole, but
there was still plenty of methane, which continued to sublimate. At
this point, the atmospheric composition was more depleted in nitrogen,
as was the ice freezing out on the winter pole on top of the earlier
deposited nitrogen rich ice. In this way, we would observe methane
becoming more diluted with increasing depth into the surface of Eris.

Atmospheric models of Pluto suggest about one meter of frost was
transported across the surface of Pluto during its last seasonal cycle
(Spencer et al., 1997).  If a similar amount of frost was transported
across the surface of Eris, our Monte Carlo ray tracing model depths
in Figure 2 are consistent with our observations probing the
atmospheric condensation of Eris during its last perihelion passage.
\bigskip


\noindent {\bf 5. Discussion \hfil}
\smallskip

Next, we compare the results reported here with results in the
literature.  In Figure 5, we plot the 729.6 nm and 889.7 nm bands
reported here (black lines) and in Licandro et al. (2006; red
lines). The most striking difference occurs between observations of
the 889.7 nm band. Not only does Licandro's 889.7 nm band have a
larger blue shift than the band reported here, but it is also deeper.

Are the differences in Figure 5 real or the result of differences in
observation or analysis techniques? Both sets of data were reduced in
the same fashion (Massey et al., 1992). The same solar analog star (SA
93-101) was used to remove telluric bands and Fraunhofer lines from
both sets of Eris spectra. Furthermore, sky line measurements indicate
the wavelength calibration of the work reported here is accurate to
0.03 nm (see Table II). The same sky line measurements indicate the
wavelength calibration of Licandro et al. is accurate to 0.1 nm.  It
is possible that incomplete subtraction of strong sky lines could be
affecting our shifts; however, removal of the strong 888.586 nm sky line
ranges from excellent (top and middle panels) to poor (bottom panel)
in Figure 2, yet the shifts of the three 889.7 nm bands are
statistically the same. Furthermore, we re-analyzed Licandro's
published Eris spectrum using the Hapke model and cross correlation
software described here. We found blue shifts of 0.08 $\pm$ 0.12 nm
and 1.38 $\pm$ 0.12 nm for the 729.6 nm and 889.7 nm bands (see Table
3 and Figure 4), i.e. we found shifts consistent with those reported
in Licandro et al. So, it appears to us that the analysis software is
not inducing the differences. We suspect the differences between the
spectra are real.

Are there any other measurements of methane band shifts in optical
spectra of Eris that support either correlation?  Alvarez-Candal et
al. (2008) reported a shift of 0.8 nm for the 729.6 nm band (i.e. a
shift quite similar to the 0.7 nm shift reported here), but a shift of
1.2 nm for the 889.7 nm band (i.e. a shift quite similar to the 1.38
nm shift calculated here from Licandro's spectrum). Their measurements
support a decrease in nitrogen with an increase in depth.

Do Eris' near-infrared methane ice bands support either correlation?
These bands are much stronger than the optical bands, so they probe
material closer to the surface than the optical bands. These bands are
blue shifted by about 1 nm; however, the uncertainty in the shifts is
about 1 nm (Brown et al., 2005; Dumas et al., 2007).  Near-infrared
spectra of Eris with ten times higher spectral resolution are
necessary to discriminate between the correlations in Figure 4.

Do other icy dwarf planets exhibit correlations between shift and
albedo? Pluto's 729.6 nm band appears to have a significantly larger
blue shift than the 889.7 nm band (Grundy and Fink, 1996). Such shifts
suggest the nitrogen abundance increases with depth. Unfortunately,
there are no measurements of other methane bands in optical spectra of
Pluto. Makemake's optical methane ice bands exhibit small blue shifts
much like Eris; however, the spectra have insufficient spectral
resolution or too few bands and a limited range of albedo to test for
a correlation (Tegler et al., 2007; Tegler et al., 2008).

It is difficult for either the atmosphere$-$surface model put forth
here or in Licandro et al. (2006) to easily explain the Eris shifts
reported here and in the literature.  Perhaps the easiest explanation
of the different observations is that they are the result of surface
heterogeneity. It is possible the observations were all taken at
significantly different longitudes. We point out that Triton is about
40${^\circ}$ from pole-on, yet it exhibits nitrogen ice bands that
vary in strength by a factor of two as it rotates on its axis (Grundy
and Young, 2004).  It seems to us that the next step is to obtain high
signal precision optical spectra of opposite hemispheres of Eris in
order to test for surface heterogeneity, and thereby help sort out
what is happening on its surface.

\noindent {\bf Acknowledgements \hfil} 
\bigskip

S.C.T., W.R., and D.C. gratefully acknowledge support from NASA
Planetary Astronomy grant NNG06G138G to Northern Arizona University
and the University of Oklahoma. W.M.G. gratefully acknowledges support
from Planetary Geology and Geophysics grant NNG04G172G to Lowell
Observatory.  We thank Steward Observatory for consistent allocation
of telescope time on the MMT.
\bigskip

\vfil\eject

\noindent{\bf References \hfil}

\noindent Alvarez-Candal, A., Fornasier, S., Barucci, M. A., de Bergh,
C., Merlin, F., 2008.  Visible spectroscopy of the new ESO large program on
trans-Neptunian objects and Centaurs. Astron. Astrophys. 487, 741-748.
\bigskip

\noindent Brown, M. E., Schaller, L., 2008. The mass of the dwarf
planet Eris. Science 316, 1585.

\noindent Brown, M. E., Trujillo, C. A., Rabinowitz, D. L.,
2005. Discovery of a planetary-sized object in the scattered Kuiper
belt. Astrophys. J. 636, L97-L100.
\bigskip

\noindent Cruikshank, D. P., Roush, T. L., Owen, T. C., Geballe, T. R., de
Bergh, C., Schmitt, B., Brown, R. H., Bartholomew, M. J., 1993. Ices
on the surface of Triton. Science 261, 742-745.
\bigskip

\noindent Dumas, C., Merlin, F., Barucci, M. A., de Bergh, C., Hainaut, O.,
Guilbert, A., Vernazza, P., Doressoundiram, A., 2007. Surface
composition of the largest dwarf planet 136199 Eris (2003
UB313). Astron. Astrophys. 471, 331-334.
\bigskip

\noindent Grundy, W. M., Buie, M. W., 2001. Distribution and evolution
of CH$_4$, N$_2$, and CO ices on Pluto's surface: 1995 to 1998. Icarus
153, 248-263.
\bigskip

\noindent Grundy, W. M., Fink, U., 1996. Synoptic CCD
spectrophotometry of Pluto over the past 15 years. Icarus 124,
329-343.
\bigskip

\noindent Grundy, W. M., Stansberry, J. A., 2000. Solar gardening
and the seasonal evolution of nitrogen ice on Triton and Pluto. Icarus
148, 340-346.
\bigskip

\noindent Grundy, W. M., Young, L. A., 2004. Near-infrared spectral
monitoring of Triton with IRTF/Spex I: establishing a baseline for
rotational variability. Icarus 172, 455-465.
\bigskip

\noindent Grundy, W. M., Schmitt, B., Quirico, E., 2002. The
temperature-dependent spectrum of methane ice I between 0.7 and 5
$\mu$m and opportunities for near-infrared remote thermometry. Icarus
155, 486-496.
\bigskip

\noindent Hanuschik, R. W., 2003. A flux-calibrated, high-resolution,
atlas of optical sky emission lines from UVES. Astron. Astrophys. 407,
1157-1164.
\bigskip

\noindent Hapke, B., 1993. Theory of Reflectance and Emittance
Spectroscopy.  Cambridge Univ. Press, New York.
\bigskip

\noindent Licandro, J., Grundy, W. M., Pinilla-Alonso, N., Leisy, P.,
2006. Visible spectroscopy of 2003 UB313: evidence for N$_2$ ice on
the surface of the largest TNO? Astron. Astrophys. 458, L5-L8.
\bigskip

\noindent Massey, P., Valdes, F., Barnes, J. 1992. A user's guide to
reducing slit spectra with IRAF. NOAO,
Tucson. http://iraf.noao.edu/iraf/ftp/iraf/docs/spect.ps.Z.
\bigskip

\noindent Owen, T. C., Roush, T. L., Cruikshank, D. P., Elliot, J. L., Young,
L. A., de Bergh, C., Schmitt, B., Geballe, T. R., Brown, R. H.,
Bartholomew, M. J., 1993. Surface ices and the atmospheric
composition of Pluto. Science 261, 745-748.
\bigskip

\noindent Quirico, E., Schmitt, B., 1997. Near-infrared spectroscopy
of simple hydrocarbons and carbon oxides diluted in solid N$_2$ and as
pure ices: Implications for Triton and Pluto. Icarus 127, 354-378.
\bigskip

\noindent Spencer, J. R., Stansberry, J. A., Trafton, L. M., Young,
E. F., Binzel, R. P., Croft, S. K., 1997. Volatile transport, seasonal
cycles, and atmospheric dynamics on Pluto. In: Stern, S. A., Tholen,
D. J, (Eds.), Pluto and Charon, Univ. Arizona Press, Tucson,
pp. 435-473.
\bigskip

\noindent Stansberry, J., Grundy, W., Brown, M., Cruikshank, D.,
Spencer, J., Trilling, D., Margot, J., 2008. Physical properties of
Kuiper belt and Centaur objects: Constraints from Spitzer Space
Telescope. In: Barucci, A., Boehnhardt, H., Cruikshank, D.,
Morbidelli, A., (Eds.), The Solar System beyond Neptune, Univ. Arizona
Press, Tucson, pp. 161-179.
\bigskip

\noindent Tegler, S. C., Grundy, W. M., Romanishin, W., Consolmagno,
G. J., Mogren, K., Vilas, F., 2007. Optical spectroscopy of the large
Kuiper belt objects 136472 (2005 FY9) and 136108 (2003
EL61). Astron. J. 133, 526-530.
\bigskip

\noindent Tegler, S. C., Grundy, W. M., Vilas, F., Romanishin, W.,
Cornelison, D., Consolmagno, G. J., 2008. Evidence of N$_2$-ice on
the surface of the icy dwarf Planet 136472 (2005 FY9). Icarus 195,
844-850.

\vfil\eject

\begin{center}
\noindent\begin{tabular}{cccccc}
\multicolumn{6}{c}{\bf Table 1 } \\ \multicolumn{6}{c}{\bf Eris Observations}\\
\hline
\\
UT Date & No. Exp.$^a$ & Tot Exp & UT Range & Airmass Range  & Wavelength \\
        &          & (hh:mm)    &  (hh:mm)      &                & (nm)    \\
\\
\hline 
\\ 
2007 Sep 08 & 14 & 02:20 & 07:49 $-$ 11:29 & 1.48 $-$ 1.25 $-$ 1.37 & 660.2 $-$ 853.0\\
2007 Sep 09 & 14 & 02:20 & 08:26 $-$ 11:33 & 1.34 $-$ 1.25 $-$ 1.39 & 758.7 $-$ 951.3 \\
2007 Sep 10 & 18 & 03:00 & 07:42 $-$ 11:32 & 1.47 $-$ 1.25 $-$ 1.40 & 758.7 $-$ 951.3 \\
2007 Sep 11 & 17 & 02:50 & 07:39 $-$ 11:32 & 1.47 $-$ 1.25 $-$ 1.41 & 758.7 $-$ 951.3 \\
\\
\hline 
\end{tabular}
\end{center}
\indent {$^a$ Number of 600 sec exposures. \hfil} \\

\vfil\eject

\begin{center}
\noindent\begin{tabular}{ccccccccc}
\multicolumn{9}{c}{\bf Table 2 } \\ \multicolumn{9}{c}{\bf Night Sky Lines}\\
\hline
\\

Sep08$^a$ & Sep09$^a$ & Sep10$^a$ & Sep11$^a$ & VLT$^b$ & Sep08$^c$ & Sep09$^c$ & Sep10$^c$ & Sep11$^c$  \\
\\
\hline 
\\ 
692.326 & & & & 692.319 & $+$0.007 & & &  \\
731.620 & & & & 731.629 & $-$0.009 & & &  \\
732.953 & & & & 732.916 & $+$0.037 & & &  \\
734.101 & & & & 734.090 & $+$0.011 & & &  \\
757.188 & & & & 757.175 & $+$0.013 & & &  \\
779.411 & 779.357 & 779.355 & 779.361 & 779.412 & $-$0.001 & $-$0.055 & $-$0.057 & $-$0.051 \\
782.161 & 782.117 & 782.119 & 782.122 & 782.152 & $+$0.009 & $-$0.035 & $-$0.033 & $-$0.030 \\
799.361 & 799.363 & 799.360 & 799.364 & 799.333 & $+$0.028 & $+$0.030 & $+$0.027 & $+$0.031 \\
839.893 & 839.880 & 839.887 & 839.881 & 839.918 & $-$0.025 & $-$0.038 & $-$0.031 & $-$0.037 \\ 
        & 886.731 & 886.683 & 886.672 & 886.761 &          & $-$0.030 & $-$0.078 & $-$0.089 \\
        & 888.580 & 888.530 & 888.535 & 888.586 &          & $-$0.006 & $-$0.056 & $-$0.051 \\
        & 890.323 & 890.270 & 890.278 & 890.312 &          & $+$0.011 & $-$0.042 & $-$0.034 \\
        & 891.966 & 891.917 & 891.921 & 891.964 &          & $+$0.002 & $-$0.047 & $-$0.043 \\
        & 894.358 & 894.308 & 894.308 & 894.341 &          & $+$0.017 & $-$0.033 & $-$0.033 \\
        & 898.887 & 898.855 & 898.838 & 898.838 &          & $+$0.049 & $+$0.017 & $+$0.000 \\
        & 903.842 & 903.808 & 903.803 & 903.806 &          & $+$0.036 & $+$0.002 & $-$0.003 \\
        &         &         &         &         &          &          &          &          \\
        &         &         &         & Avg Dif & $+$0.008 & $-$0.002 & $-$0.030 & $-$0.031 \\
        &         &         &         & Std Dev & $+$0.018 & $+$0.034 & $+$0.033 & $+$0.032 \\
\\  
\hline 
\end{tabular}
\end{center}
\indent{$^a$ MMT values from our spectra in nm. \hfil} \\
\noindent{$^b$ VLT values from Hanuschik (2003) in nm. \hfil} \\
\noindent {$^c$ Difference between MMT and VLT sky line measurements in nm. \hfil} \\

\vfil\eject

\begin{center}
\noindent\begin{tabular}{lccccc}
\multicolumn{6}{c}{\bf Table 3 } \\ \multicolumn{6}{c}{\bf Blue Shifts of Methane Ice Bands$^a$}\\
\hline
\\
Band & Wavelength & Blue Shift$^a$ & Albedo$^a$ & Blue Shift$^b$ & Albedo$^b$ \\
     &  (nm)      &  (nm )     &     & (nm) &    \\
\\
\hline 
\\ 
3$\nu_1$ $+$ 4$\nu_4$ & 729.6 & 0.70 $\pm$ 0.11 & 0.79 & 0.08 $\pm$ 0.12 & 0.77 \\
3$\nu_3$ $+$ 2$\nu_4$ & 869.1 & 0.60 $\pm$ 0.10 & 0.79 &                  &      \\
2$\nu_1$ $+$ $\nu_3$ $+$ 2$\nu_4$& 889.7 & 0.46 $\pm$ 0.18 & 0.52 & 1.38 $\pm$ 0.12 & 0.31\\
3$\nu_1$ $+$ 2$\nu_4$ & 896.8 & 0.56 $\pm$ 0.13 & 0.69     &  &\\
2$\nu_3$ $+$ 4$\nu_4$ & 901.9 & 0.56 $\pm$ 0.13 & 0.69     &  &\\
                      & 918.0 & 0.77 $\pm$ 0.16 & 0.94     &  &\\

\\
\hline 
\end{tabular}
\end{center}
\indent{$^a$ This work. Single cross correlation on 896.8 and 901.9 nm bands. \hfil} \\
\noindent{$^b$ Licandro et al. (2006). \hfil} \\

\vfil\eject

\noindent {\bf Figure Captions \hfil}
\bigskip

\noindent {\bf Fig. 1.} Reflectance spectra of Eris taken with the 6.5
meter MMT telescope on 2007 September 9, 10, and 11 UT, and a spectrum
of pure methane ice. Methane ice bands are present in the Eris
spectra.
\bigskip

\noindent {\bf Fig. 2.} Portions of the Eris spectra in Figure 1
containing the 889.7 nm methane band (black lines), and pure methane
ice Hapke models (red lines). From cross correlation experiments, we
found the 889.7 nm band is blue shifted by 0.49 $\pm$ 0.08 nm, 0.51
$\pm$ 0.10 nm, and 0.31 $\pm$ 0.23 nm relative to pure methane ice on
the nights of 2007 September 9 UT (top panel), September 10 UT (middle
panel), and September 11 UT (bottom panel). The blue shifts suggest
the presence of another ice component, probably nitrogen ice.
\bigskip

\noindent {\bf Fig. 3.}  MMT spectra of Eris (black lines) and pure
methane ice Hapke models (red lines). Starting in the upper left corner
and then clockwise are the 729.6 nm band (MMT exposure time of 2hr 20
min on 2007 September 8 UT ), 869.1 nm band, 889.7 nm band, and the
896.8 nm band with the 901.9 nm band (all with MMT exposure time of 8
hr 10 min from 2007 September 9, 10, and 11 UT).
\bigskip

\noindent {\bf Fig. 4.} Blue shift vs. albedo at maximum absorption
for five methane ice bands in MMT spectra of Eris (solid squares;
Table 3). The scale across the top of the diagram, approximate mean
depth sampled, comes from a Monte Carlo ray tracing model (Grundy and
Stansberry, 2000). The solid line is a least squares fit to the five
points.  The Spearman rank correlation coefficient for the five points
is r = 0.97.  The correlation suggests methane ice becomes more
diluted by another ice component, possibly nitrogen ice, with
increasing depth into the surface of Eris. Shifts and albedos from a
re-analysis of Licandro's spectrum (solid circles; Table
3). Licandro's measurements suggest methane ice becomes less diluted
with an increase in depth. We suspect the differences between the two
works are real and may be the result of a heterogeneous surface.

\noindent {\bf Fig. 5.} Portions of Eris reflectance spectra
containing the 729.6 nm band (top panel) and the 889.7 nm band (bottom
panel) reported here (black lines) and by Licandro et al. (red
lines). The most striking difference occurs between the 889.7 nm
bands. The band reported by Licandro not only has a larger blue shift,
but it is also deeper than the band reported here. We suspect the
differences between the two sets of observations are real.

\vfil\eject

\begin{figure}
        \centering
        \scalebox{1.25}{\includegraphics{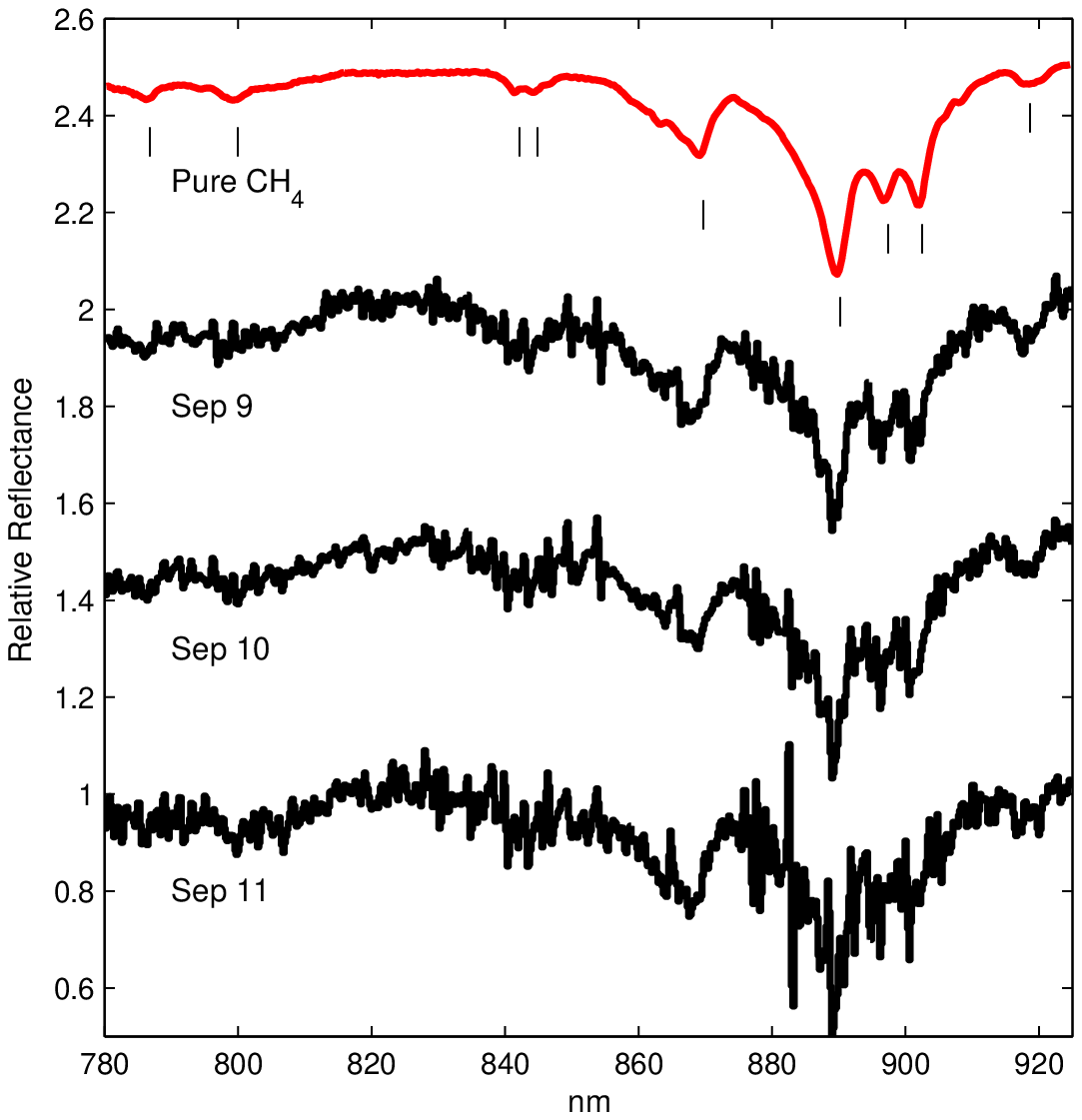}}
	\caption{}
        \end{figure}

\clearpage

\begin{figure}
        \centering
        \vskip -1.0truein
        \scalebox{1.25}{\includegraphics{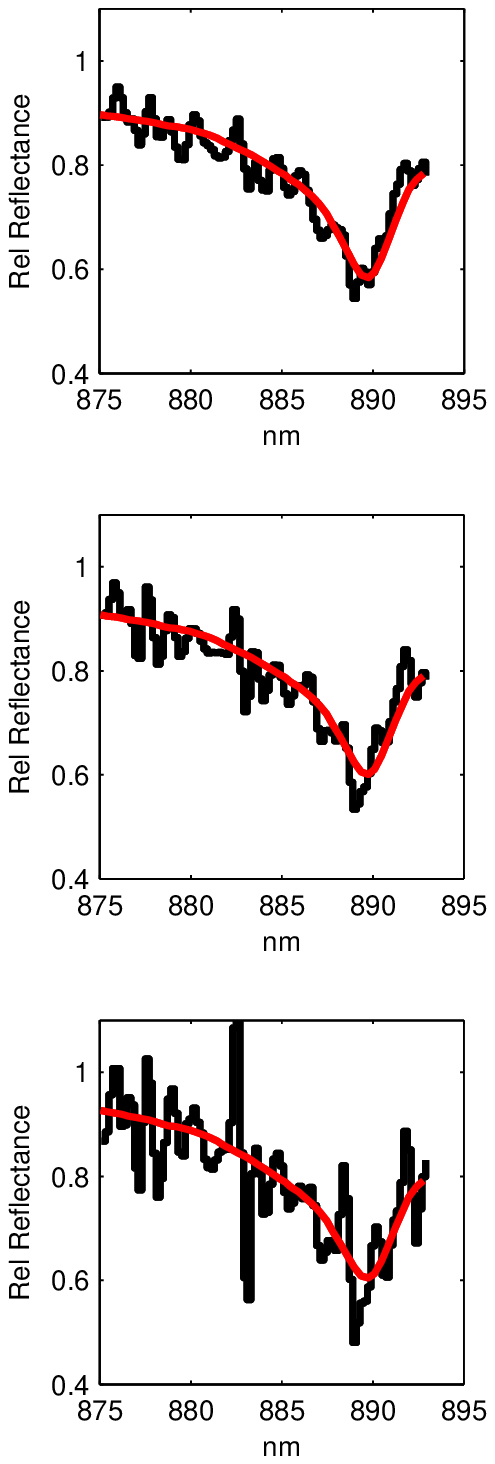}}
        \vskip -0.5truein
	\caption{}
        \end{figure}

\clearpage 

\begin{figure}
        \centering
       \scalebox{1.25}{\includegraphics{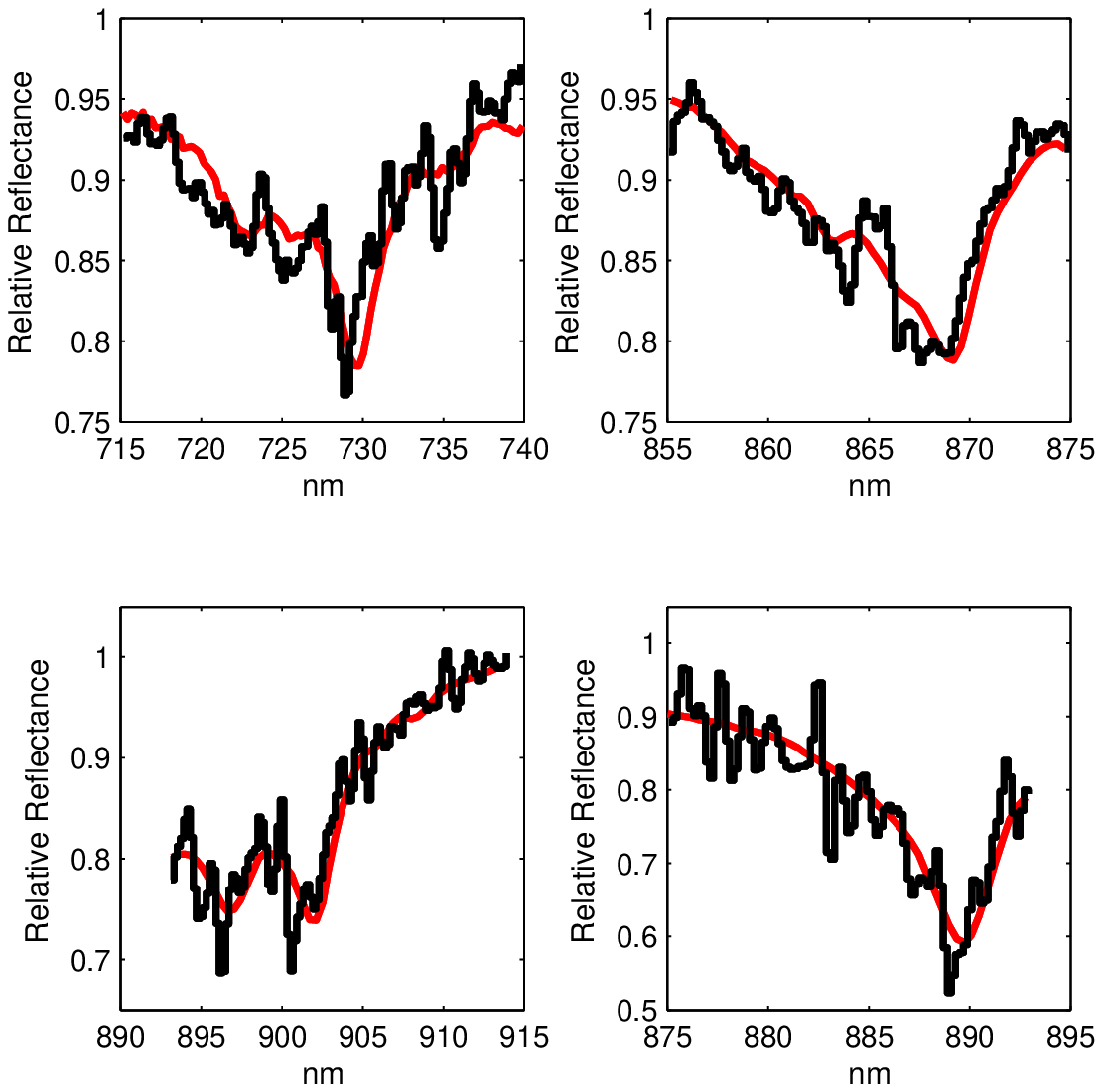}}
	\caption{}
        \end{figure}

\clearpage

\begin{figure}
	\centering
	\scalebox{0.8}{\includegraphics{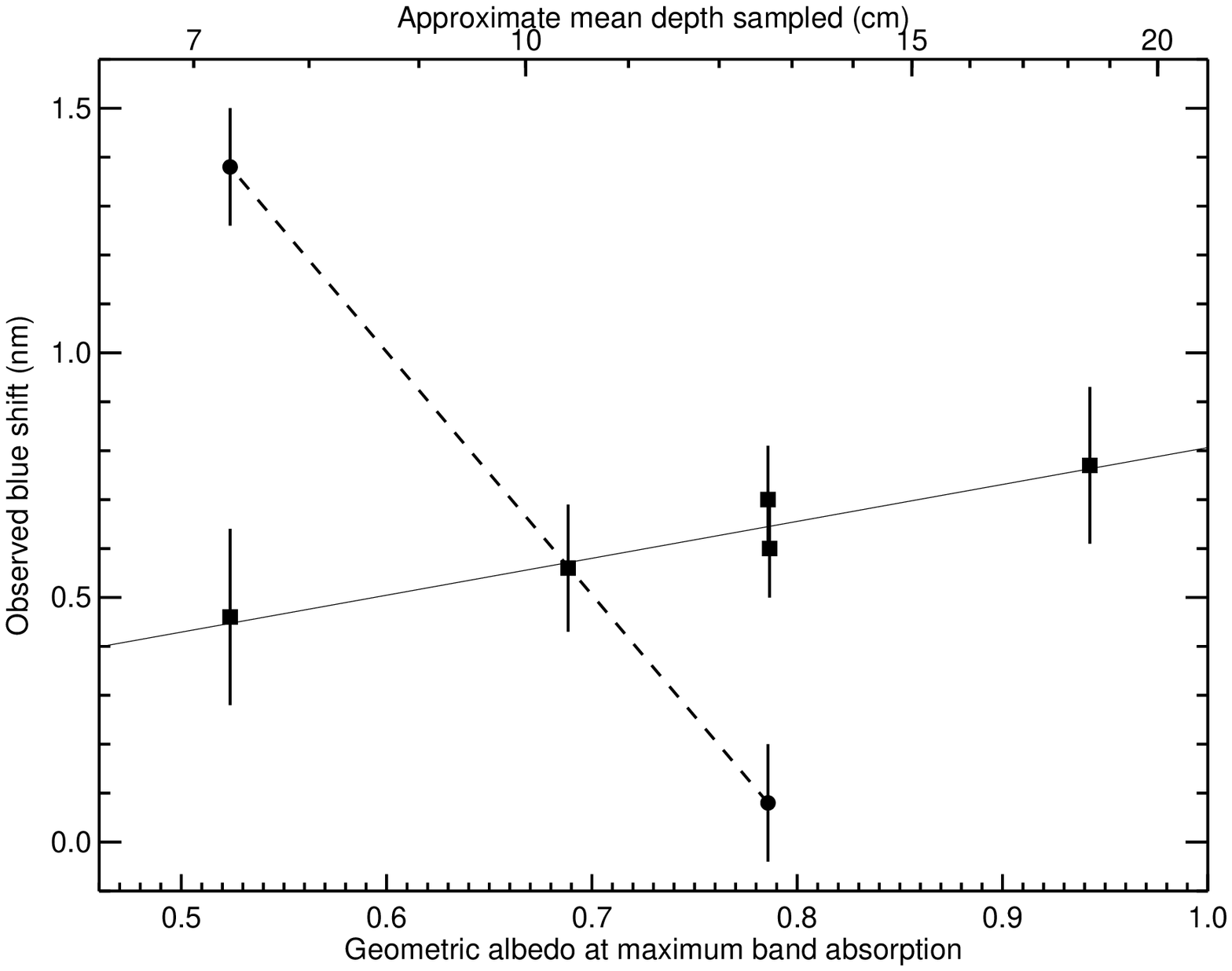}}
	\caption{}
	\end{figure}

\clearpage

\begin{figure}
	\centering
	\scalebox{1.25}{\includegraphics{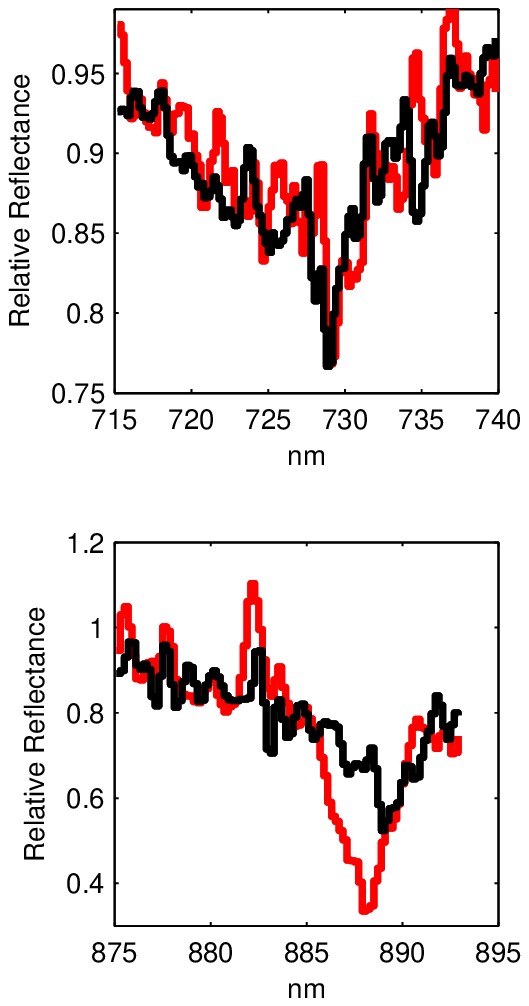}}
	\caption{}
	\end{figure}

\end{document}